\documentclass[
 reprint,
 amsmath,amssymb,
 aps
]{revtex4-2}

\usepackage{graphicx}
\usepackage{dcolumn}
\usepackage{bm}
\usepackage{bbold}
\usepackage{float}
\usepackage{braket}
\usepackage{color}

\begin{document}

\preprint{APS/123-QED}

\title{
Floquet Anderson Localization of Two Interacting Discrete Time Quantum Walks 
}

\author{Merab Malishava$^{1,2}$}
\author{Ihor Vakulchyk$^{1,2}$}%
\author{Mikhail Fistul$^{1,3}$}
\author{Sergej Flach$^1$}
\affiliation{%
\mbox{$^1$Center for Theoretical Physics of Complex Systems, Institute for Basic Science(IBS), Daejeon, Korea, 34126}\\
$^2$Basic Science Program, Korea University of Science and Technology(UST), Daejeon, Korea, 34113
 \\
$^3$Russian Quantum Center, National University of Science and Technology “MISIS”, 119049 Moscow, Russia
}
\date{\today}
\begin{abstract}
We study the interplay of two interacting discrete time quantum walks in the presence of disorder. Each walk is described by a Floquet unitary map 
defined on a chain of two-level systems. Strong disorder induces a novel Anderson localization phase with a gapless Floquet spectrum and one unique localization length $\xi_1$
for all eigenstates for noninteracting walks. We add a local contact interaction which is parametrized by a phase shift $\gamma$.
A wave packet is spreading subdiffusively beyond the bounds set by $\xi_1$ and saturates at a new length scale $\xi_2 \gg \xi_1$.
In particular we find $\xi_2 \sim \xi_1^{1.2}$ for $\gamma=\pi$. We observe a nontrivial dependence of $\xi_2$ on $\gamma$, with a maximum value
observed for $\gamma$-values which are shifted away from the expected strongest interaction case $\gamma=\pi$. The novel Anderson localization regime violates single parameter scaling for both interacting and noninteracting walks.
\end{abstract}

\maketitle

\section{\label{sec:level1}Introduction}
Anderson localization (AL) 
\cite{anderson1958absence,Lee1985,kramer1993localization,LGP} established that in the presence of uncorrelated on-site random potential, all eigenstates are exponentially localized in one and two dimensions. In three dimensions, there is an energy mobility edge separating localized and delocalized eigenstates. 
The localization length $\xi_1$ is determined by many parameters such as eigenstate energy, hopping integrals between adjacent sites, and the amplitude of the random potential, as obtained for different lattices and various types of random potentials \cite{LGP}. AL results in a strong suppression of transport in low-dimensional systems \cite{anderson1958absence,LGP}. AL was observed experimentally in a variety of condensed matter and optical systems \cite{kramer1993localization,lahini2008anderson,billy2008direct,roati2008anderson,strozer2006observation,schwartz2007transport}.   

The challenging study of the interplay of interaction and disorder leads to a number of unexpected results 
for the localization properties of many particles eigenstates.  The seemingly simplest case of two interacting particles (TIP) in one space dimension
was analyzed in an impressive set of publications
\cite{shepelyansky1994coherent,jacquod1997breit,imry1995coherent,roemer1997no,vonoppen1996interaction,song1999general,frahm1995scaling,frahm2016eigenfunction,krimer2010statistics,krimer2011two,ortuno1999localized,Sergej_ivanchenko_2014,sergej_ivanchenko_2017}. 
For uncorrelated disorder the TIP localization length $\xi_2$ is assumed to be finite, with the main questions addressing the way $\xi_2$ scales with $\xi_1$ in the 
limit of weak disorder
\cite{shepelyansky1994coherent,jacquod1997breit,imry1995coherent,roemer1997no,vonoppen1996interaction,song1999general,frahm1995scaling,frahm2016eigenfunction,krimer2010statistics,krimer2011two,ortuno1999localized}, and the nature of the observed sub-diffusive wave packet spreading on length scales $\xi_1 \ll L \ll \xi_2$ \cite{Sergej_ivanchenko_2014,sergej_ivanchenko_2017}.
Lack of analytical results stresses the need of computational studies. However, in all above cases, there are limits set by the size of the system (in particular
for diagonalization routines due to immense Hilbert space dimensions), the largest evolution
times obtained through direct integrations of time-dependent Schr\"odinger equations with continuous time variables, and the energy dependence of the localization length
$\xi_1$.

An interesting alternative platform is Floquet unitary maps on two-level system networks known as
\textit{discrete-time quantum walks} (DTQW). 
DTQWs were introduced for quantum computing purposes \cite{aharonov2001quantum,PhysRevA.48.1687,doi:10.1080/00107151031000110776,tregenna2003controlling}. Recently they have been used to study some numerically challenging complex problems of condensed matter physics, e.g. lattice Dirac transport \cite{chandrashekar2013two}, topological phases \cite{obuse2011topological}, Anderson localization 
\cite{PhysRevA.94.023601,vakulchyk2017anderson}, and nonlinear transport in ordered and disordered lattices supporting flat bands \cite{vakulchyk2019wave,vakulchyk2018almost}. 
Note that the resulting Floquet Anderson localization is proven analytically for a whole range of different cases, including those where the eigenvalue spectrum
is dense, homogeneous and gapless, and the localization length $\xi_1$ is governing all random eigenstates  independent of their eigenvalues 
\cite{vakulchyk2017anderson}.
We stress that DTQWs are particular examples of a Floquet driven quantum lattice, and, therefore, 
apply to studies of AL under non-equilibrium or simply infinite temperature conditions in an elegant and simple way as compared to the approach defined
by time-periodic Hamiltonian systems (see e.g. \cite{Ducatez_2017}).
DTQWs have been implemented in various condensed matter and optics setups, see e.g. \cite{CoinOperImpl,Impl1,di2004cavity,Impl3}. 

Stefanak et al \cite{stefanak_2011} and Ahlbrecht et al \cite{Ahlbrecht_2012} proposed an extension of the single particle DTQW to two interacting DTQWs using a local contact interaction which is parametrized by a phase shift $\gamma$. We use this Hubbard-like interaction and consider two interacting disordered
discrete time quantum walks (TIW).
Using direct numerical simulations, we compute the time-dependent spreading of the TIW wave packet and its dependence on the angle $\xi_1$, and the strength of the interaction $\gamma$. 
The computational evolution of wave functions for Hamiltonian systems involves the need to control accumulating errors due to the discretization of the continuous time variable.
DTQWs do not require such approximations making them superior when it comes to long time evolutions.

The paper is organized as follows: in Sec. II, we first present the model for a single one-dimensional DTQW. We then extend the model to two interacting DTQWs.
In Sec. III, we present the computational details and measures used for the study of the time evolution of two interacting DTQWs. 
In Sec. IV we present the numerical results, and discuss them.
Sec. V provides the conclusions.

\section{\label{sec:level2}
Models}

We consider the dynamics of a single quantum particle with an internal spin-like degree of freedom on a one-dimensional lattice \cite{aharonov2001quantum,PhysRevA.48.1687,doi:10.1080/00107151031000110776,tregenna2003controlling,obuse2011topological,Venegas-Andraca2012,chandrashekar2008optimizing,vakulchyk2017anderson}. 
Such a system is characterized by a two-component wave function   $\ket{\Psi(t)}$  defined on a discrete chain of $N$ sites. The wave function is
embedded in a $2N$-dimensional Hilbert space:
\begin{eqnarray}\label{single_particle_wf}
    \ket{\Psi(t)} =\sum_{n=1}^{N} \sum_{\alpha=\pm} \psi_n^{\alpha}(t) \ket{\alpha} \otimes \ket{n} = \nonumber \\ \sum_{n=1}^{N} \Big[\psi_n^+(t)\ket{+} + \psi_n^-(t) \ket{-}\Big] \otimes \ket{n},
\end{eqnarray}
where $\ket{\alpha}=\ket{\pm}$ are basis vectors of {\it local two-level systems}, $\ket{n}$ are basis vectors in a {\it one-dimensional coordinate space}, and $\psi_n^{\alpha}$ are the wave function amplitudes.
The Floquet time evolution of the system is realized by means of a unitary map involving coin $\hat C$ and shift $\hat S$ operators:
\begin{eqnarray}
\label{evolution}
    \ket{\Psi(t+1)} = \hat S \hat C \ket{\Psi(t)}.
\end{eqnarray}
The coin operator $\hat C$ is a unitary matrix given by
\cite{vakulchyk2017anderson}
\begin{eqnarray}\label{Coin1}
    \hat{C}=\sum_{n=1}^N
    \hat{c}_n
    \otimes\ket{n}\bra{n}
\end{eqnarray}
with local unitary coin operators $\hat{c}_n$ 
\begin{eqnarray}\label{Coin12}
\hat{c}_n=
e^{i\varphi_n}
\begin{pmatrix}
        e^{i\varphi_{1,n}}\cos\theta_n   & e^{i\varphi_{2,n}}\sin\theta_n \\
        -e^{-i\varphi_{2,n}}\sin\theta_n & e^{-i\varphi_{1,n}}\cos\theta_n
    \end{pmatrix} 
\end{eqnarray}
which 
are parametrized by four spatially dependent angles $\theta_n$, $\varphi_n$, $\varphi_{1,n}$ and $\varphi_{2,n}$. Such local coin operators can be implemented in various experimental setups through e.g. a periodic sequence of effective magnetic field pulses \cite{CoinOperImpl,Impl1,di2004cavity,Impl3}. 

As it was shown in Ref. \cite{vakulchyk2017anderson}, the angle $\phi_n$ is related to a potential energy, the angles $\phi_{1,n}$ and $\phi_{2,n}$ to an external and internal
magnetic flux respectively, and the angle $\theta_n$ to a local kinetic energy or hopping. In this work we intend to generalize the corresponding problem of two
interacting particles in a one-dimensional tight-binding chain with uncorrelated disorder. Therefore we choose $\phi_{1,n}=\phi_{2,n}=0$
and $\theta_n \equiv \theta$, which simplifies the local coins $\hat{c}_n$ in (\ref{Coin12}) to
\begin{eqnarray}\label{Coin13}
\hat{c}_n=
e^{i\varphi_n}
\begin{pmatrix}
        \cos\theta   & \sin\theta \\
        -\sin\theta & \cos\theta
    \end{pmatrix} \;.
\end{eqnarray}
The spatial local disorder will be introduced through the angles $\varphi_n$ \cite{vakulchyk2017anderson}. This particular choice of disorder resembles a random on-site potential of the  original Anderson model \cite{anderson1958absence}.    

The shift operator $\hat{S}$ in Eq. (\ref{evolution}) 
couples neighboring sites by shifting all the $\psi_n^+$ components one step to the right, and all the $\psi_n^-$ components to the left: 
\begin{eqnarray}\label{Shift1}
    \hat{S} = \sum_n \ket{n}\bra{n+1} \otimes \ket{-}\bra{-} \; +\; \ket{n}\bra{n-1} \otimes \ket{+}\bra{+}. \nonumber\\ 
\end{eqnarray}
This completes the definition of
a single particle discrete-time quantum walk 
\cite{aharonov2001quantum,PhysRevA.48.1687,doi:10.1080/00107151031000110776,tregenna2003controlling,obuse2011topological,Venegas-Andraca2012,chandrashekar2008optimizing,vakulchyk2017anderson}.

\begin{figure}
\includegraphics[width=0.5\textwidth]{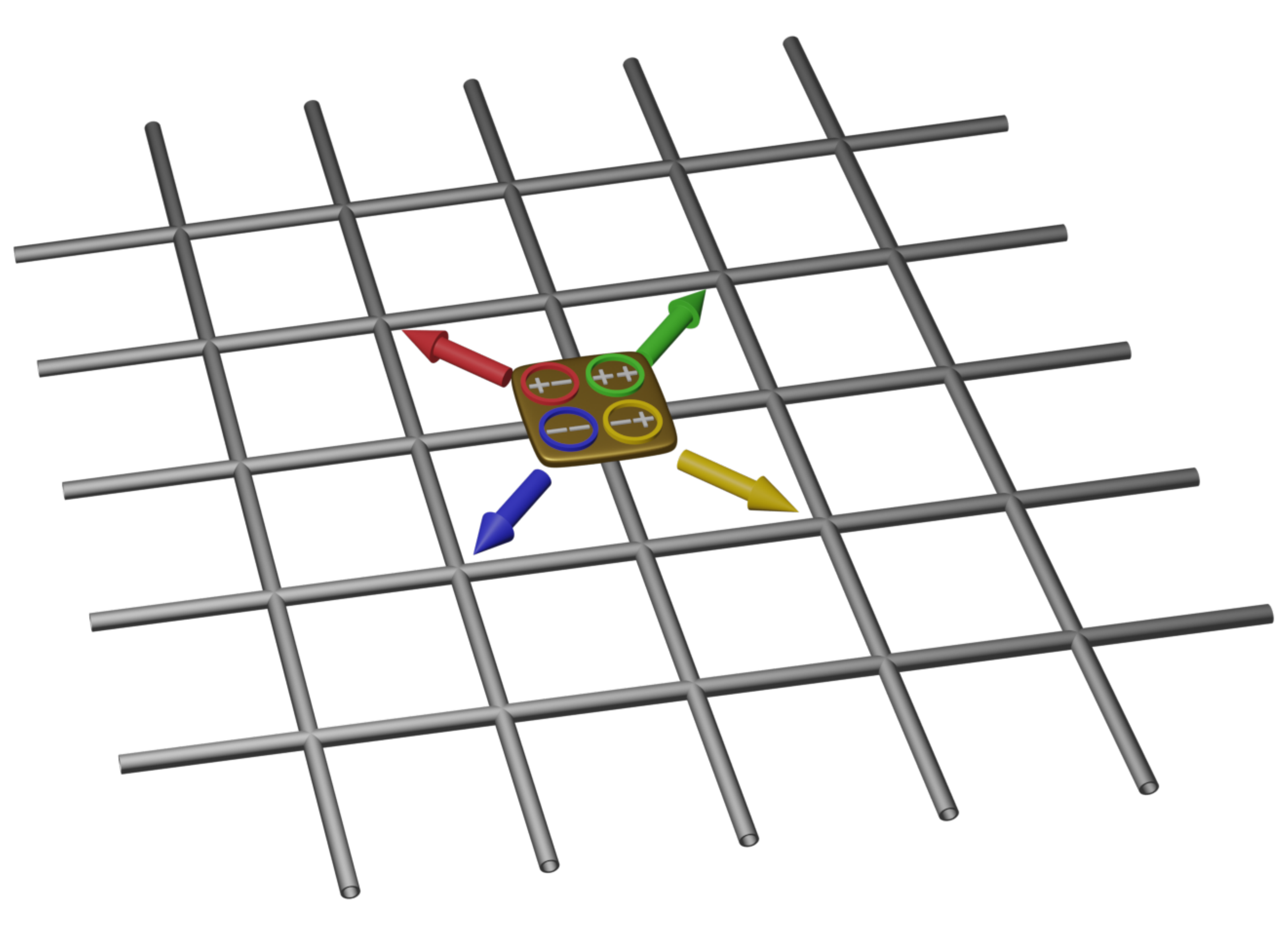}
\caption{\label{Scheme} 
A schematic view of the TIW. The four components of the wave function on each site of a square lattice are shifted in different directions indicated by the arrows.
}
\end{figure}

We extend the above single particle walk to two interacting discrete time quantum walks (TIW) in analogy to the extension of a single quantum particle in an Anderson model
to two interacting particles:
\begin{eqnarray}\label{two_particle_wave_function}
    \ket{\Psi(t)} =
\sum_{i,j=1}^N\sum_{\alpha,\beta = \pm} \psi_{ij}^{\alpha\beta}(t)\ket{\alpha,\beta} \otimes \ket{i,j}.
\end{eqnarray}
The wave function $\ket{\Psi(t)}$ is embedded in a $4N^2$-dimensional Hilbert space where $\ket{\alpha,\beta}$ are basis vectors of two local two-level systems,
and $\ket{i,j}$ are basis vectors in a two-dimensional square lattice.
The TIW evolution is a obtained through a product of a TIW coin $\hat{W}$, shift $\hat T$ and interaction $\hat G$ operators acting on the wave function
(see Fig.\ref{Scheme}):
\begin{eqnarray}\label{tip_evolution}
    \ket{\Psi(t+1)} = \hat T \hat W \hat G \ket{\Psi(t)}.
\end{eqnarray}
 The coin $\hat{W}$ and shift $\hat{T}$ operators are tensor products of the corresponding single particle operators:
\begin{equation}
 \hat{W} = \hat{C}\otimes\hat{C} \;,\; \hat{T} = \hat{S}\otimes\hat{S}\;.
\end{equation}
In the absence of an interaction $\hat{G} = \mathbb{1}$ they describe the
evolution of two independent single particle DTQWs. 
The local Hubbard-like contact interaction between the two DTQWs was 
introduced in Ref. \cite{Ahlbrecht_2012} as
\begin{eqnarray}\label{tip_interaction}
    \hat{G} = \mathbb{1}_c \otimes \mathbb{1}_p +\left(
e^{i\gamma} -1\right) \mathbb{1}_c \otimes \hat N , 
\end{eqnarray}
where $\gamma$ is the interaction strength parameter. 
$\hat{N}=\sum_i\ket{i,i}\bra{i,i}$ is a projector on the diagonal of the coordinate space, 
$ \mathbb{1}_c$ is the $4 \times 4$ unity matrix in the coin space, and $ \mathbb{1}_p$ is the $N^2 \times N^2$ unity matrix in the position
space. Note that 
$\gamma=0$ corresponds to two noninteracting DTQWs. 

\section{Anderson localization}

The local disorder is introduced through uncorrelated random values of the angle  $\varphi_n$. 
For the disorder strength $0 \leq W \leq 2\pi$, a set of $\varphi_n$ is independently drawn from a uniform distribution of $[-W/2, W/2]$.

\subsection{Single particle DTQW}

As it was shown in Ref. \cite{vakulchyk2017anderson}, all eigenstates of the single particle DTQW are exponentially localized and characterized by a localization length
$\xi_1$, in full analogy to Anderson localization for Hamiltonian single particle systems \cite{anderson1958absence}.
The single particle DTQW possesses two distinct limiting parameter cases for which $\xi_1 \rightarrow \infty$. The first is obtained for $W \rightarrow 0$, again in full
analogy with Hamiltonian systems. The DTQW eigenvalues form a band spectrum and are located on the unit circle \cite{vakulchyk2017anderson}, which is in general gapped
for $W \rightarrow 0$. Consequently the localization length $\xi_1$ is a function of the eigenvalue and different for different eigenstates, reaching its largest value in the center of the above bands.
The second parameter case is unique for Floquet Anderson systems and is obtained for the case of {\it strongest} disorder $W=\pi$.
The DTQW spectrum is now dense, homogeneous and gapless on the unit circle, with all eigenstates having the same localization length irrespective of their
eigenvalue \cite{vakulchyk2017anderson}:
\begin{eqnarray}
    \xi_1 = -\frac{1}{\ln \left( |\cos\theta| \right)}. 
\label{loc_length_analyt}
\end{eqnarray}
The limit $\xi_1 \rightarrow \infty$ is obtained by varying the hopping angle $\theta \rightarrow 0$. We are not aware of a similar regime for
Hamiltonian systems. In the following, we will study the TIW in that novel regime.

\subsection{TIW}

We will follow the time evolution of a TIW wave function 
starting from the initial state 
\begin{eqnarray}
    \ket{\Psi(t=0)} = \frac{(\ket{+,-}+\ket{-,+})}{\sqrt{2}}\otimes\ket{N/2,N/2}
\end{eqnarray}
for which the two single particle DTQWs are localized on the lattice site $N/2$ where the TIW interaction is present.
The system size $N$ varies from $5000$ up to $25000$, such that the spreading wave packet does not reach the edges in order to exclude finite system size corrections.
We perform a direct numerical propagation of \eqref{tip_evolution} up to $t_\text{max}$ which varies from $10^4$ for the $\gamma=0$ to $10^6$ for nonzero interaction
strength values. 

We follow the wave function probability distribution in coordinate space
\begin{equation}\label{prob_distr_2d}
    p_{ij}(t) = \sum_{\alpha,\beta=\pm} \left| \psi_{ij}^{\alpha\beta} \right|^2.
\end{equation}
To assess TIW localization length scales we will project $p_{ij}$ in three different ways onto a one-dimensional coordinate space and compute the
standard deviation of a probability distribution vector $\{v_i\}$ (see e.g. \cite{Sergej_ivanchenko_2014, sergej_ivanchenko_2017})
\begin{eqnarray}\label{generic_sd}
    \sigma\left[\{v_i\}\right] = \left(\sum_i i^2 v_i - \left( \sum_i i v_i \right)^2 \right)^{1/2}.
\end{eqnarray}

\textit{Measure 1: projection on a one particle space}: we define $v_i = \sum_j p_{ij}(t)$, substitute in \eqref{generic_sd} and obtain $\sigma_{1}(t)$. 

\textit{Measure 2: projection on the space of the center mass motion}: we define $v_i =  \sum_j p_{i,j-i}(t)$, substitute in \eqref{generic_sd} and obtain 
$\sigma_{\parallel}(t)$.

\textit{Measure 3: projection on the space of (relative) distance between particles}: we define $v_i = \sum_j p_{i,i+j}(t) $,  and substitute in \eqref{generic_sd} and obtain 
$\sigma_{\perp}$.

In addition to the above three TIW length scales $\sigma_1,\sigma_{\parallel},\sigma_{\perp}$ we also define a length scale $\sigma_{sp}$ which
follows from the numerical simulation of a single particle DTQW. We define 
$v_i = |\psi_i^+ (t)|^2 + |\psi_i^- (t)|^2$, substitute in \eqref{generic_sd} and obtain $\sigma_{sp}$. 

In the presence of Anderson localization, all the above length scales are expected to grow in time and saturate at some finite values for $t \rightarrow \infty$.
For the single particle DTQW we expect $\sigma_{sp}(t\rightarrow \infty) \sim \xi_1$. 
For the noninteracting TIW case $\gamma=0$ we expect the distribution $p_{ij}(t \rightarrow \infty)$ to have four-fold discrete rotational symmetry (see e.g. inset
in Fig.\ref{fig1}).
It follows $\sigma_1\approx\sigma_{\parallel}\approx\sigma_{\perp} \approx \sigma_{sp} \sim \xi_1$.
However, for $\gamma \neq 0$ the two walks are expected to be able to travel beyond the limits set by $\sigma_{sp}$ and $\xi_1$ as long as their
two coordinates are close enough such that $|i-j| < \xi_1$. This is in analogy to two interacting particles in Hamiltonian settings. The interaction is introducing
nonzero matrix elements between the Anderson eigenstates of the noninteracting system which leads to an effective internal degree of freedom of two walks (or particles)
which form a weakly bound state. Consequently the  distribution $p_{ij}(t \rightarrow \infty)$ should elongate along the diagonal $i=j$
and reduce its symmetry to a two-fold rotational symmetry (see e.g. inset 
in Fig.\ref{fig3}). It follows $\sigma_1 \approx \sigma_{\parallel} \equiv \xi_2$, $\sigma_{\perp} \approx \sigma_{sp} \sim \xi_1$ and $\xi_2 \gg \xi_1$. The TIW is 
therefore characterized by two length scales $\xi_2$ and $\xi_1$.

\section{Computational results}
\subsection{$\gamma=0$}
\begin{figure}
\includegraphics[width=0.5\textwidth]{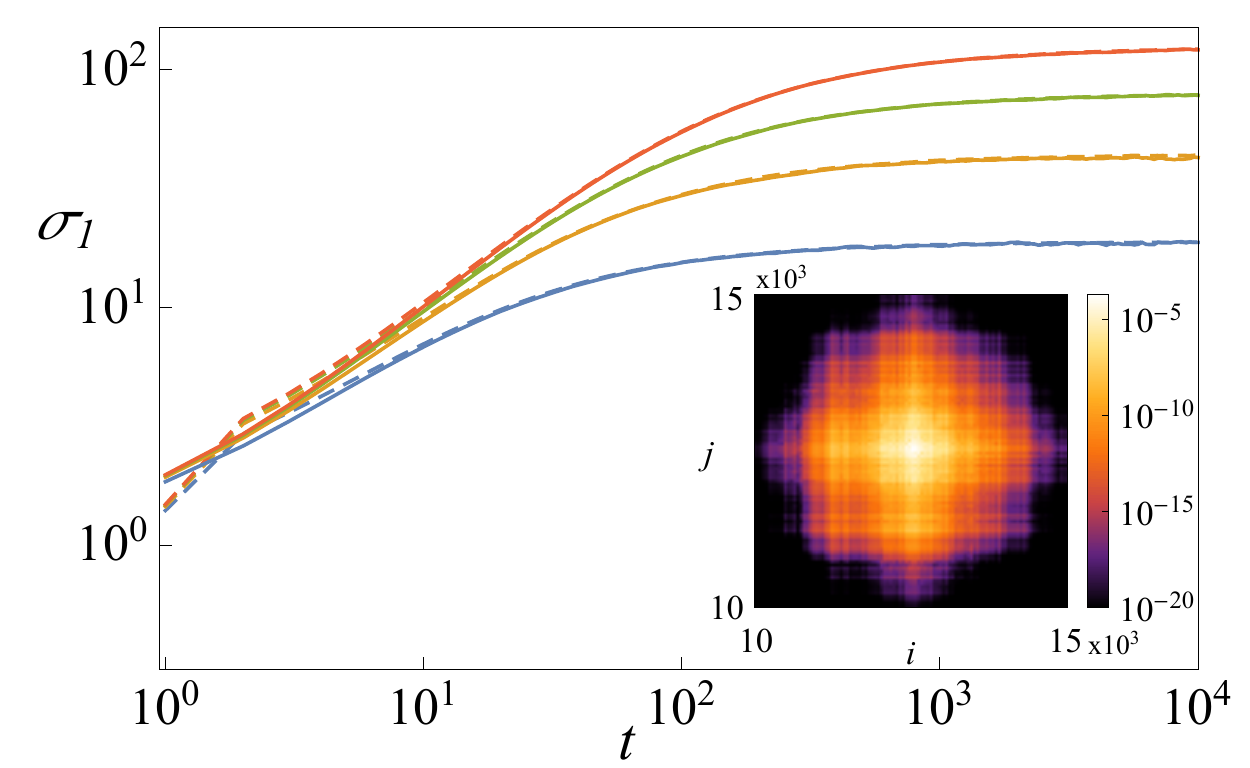}
\caption{\label{fig1} 
$\sigma_1(t)$ for $\gamma=0$ (solid lines, 100 disorder realizations) and $\sigma_{sp}(t)$ (dashed lines, average over $10^4$ disorder realizations). $\theta = \pi/8,\pi/12,\pi/16,\pi/20$ from bottom to top. 
Here $N=5000$ for $\theta=\pi/8,\pi/12$ and $N=25000$ for $\theta=\pi/16,\pi/20$.
Inset: snapshot of the probability distribution $p_{ij}(t=10^4)$ 
for $\theta=\pi/20$. 
}
\end{figure}

The time dependence $\sigma_1(t)$ (averaged over 100 disorder realizations) is shown in Fig.\ref{fig1} for various values of the hopping angle $\theta$ with solid lines. 
We observe the expected saturation of $\sigma_1$ for large evolution times $t=10^4$.
The wave function probability distribution $p_{ij}(t=10^4)$  is shown in the inset of Fig.\ref{fig1} for $\theta=\pi/20$. It shows the above discussed
four-fold discrete rotational symmetry. In addition we plot the time dependence of $\sigma_{sp}(t)$ with dashed lines, which are averaged over
$10^4$ disorder realizations and nicely follow the 
corresponding $\sigma_1(t)$ curves.

\begin{figure}
\includegraphics[width=0.5\textwidth]{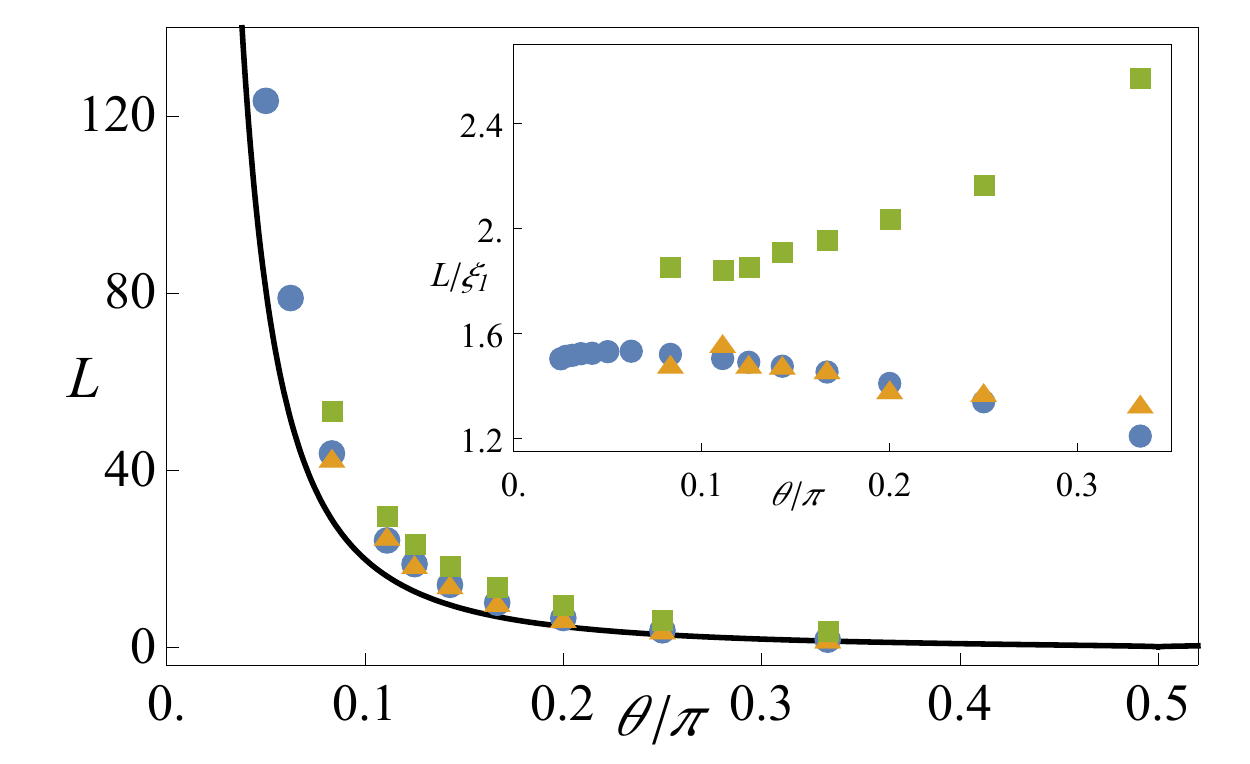}
\caption{\label{fig2}
Various length scales $L$ versus $\theta$: $\xi_1$ (solid line), $\sigma_{sp}(t_f)$ (blue circles), $\sigma_1(t_f)$ (red triangles), $P$ (green squares). Here $t_f=2\cdot 10^4$, $\gamma=0$.
Inset: $\sigma_{sp}/\xi_1$, $\sigma_1/\xi_1$ and $P/\xi_1$ as a function of the angle $\theta$. 
}
\end{figure}

A first nontrivial test is the comparison of $\xi_1$ with $\sigma_{sp}(t \rightarrow \infty)$ and $\sigma_1(t \rightarrow \infty)$ for $\gamma=0$. While we expect
$\sigma_{sp} \approx \sigma_1$, the connection between $\xi_1$ and $\sigma_{sp}$ is far from obvious. The Hamiltonian case is known to obey 
the single parameter scaling property \cite{mackinnon1983scaling}, which implies in our case $\xi_1 \sim \sigma_{sp}$. 
In Fig.\ref{fig2} we compare the localization length $\xi_{1}$ \eqref{loc_length_analyt} (solid line) with $\sigma_{sp}(t=10^4)$ from Fig.\ref{fig1} (blue circles) and 
$\sigma_{1}(t=10^4)$ from Fig.\ref{fig1} (red triangles) for different values of $\theta$. At a first glance the single parameter scaling seems to be satisfied, since the data symbols follow the analytical curve reasonably closely.
However, the inset in Fig.\ref{fig2} plots the corresponding ratios $\sigma_{sp}/\xi_1$ and $\sigma_1/\xi_1$ versus $\theta$ which result in non-horizontal curves
and indicate a violation of the single parameter scaling hypothesis. To independently confirm the absence of the single parameter scaling property, we 
diagonalize the single particle DTQW numerically for a system size $N=1500$, and obtain the participation numbers $P_{\nu}$ of all eigenfunctions $\ket{\Psi}_{\nu}$ as 
$1/P=\sum_{n=1}^{N} \sum_{\alpha=\pm} |\psi_n^{\alpha}|^4$. The average $P=\sum_{\nu} P_{\nu}/{2N}$ is plotted in Fig.\ref{fig2} (green squares). We find that $P/\xi_1$ is varying with $\xi_1$, and even shows
an opposite trend as compared to $\sigma_{sp}/\xi_1$, confirming the presence of a variety of different length scales in the problem.

At the same time, the ratio $\sigma_{sp}$ closely follows $\sigma_1$, implying that any changes in $\sigma_1$ upon increasing
the TIW interaction $\gamma$ away from $\gamma=0$ are solely due to the interaction, and not due to measurement ambiguities.

\subsection{$\gamma = \pi$}

\begin{figure}
\includegraphics[width=0.5\textwidth]{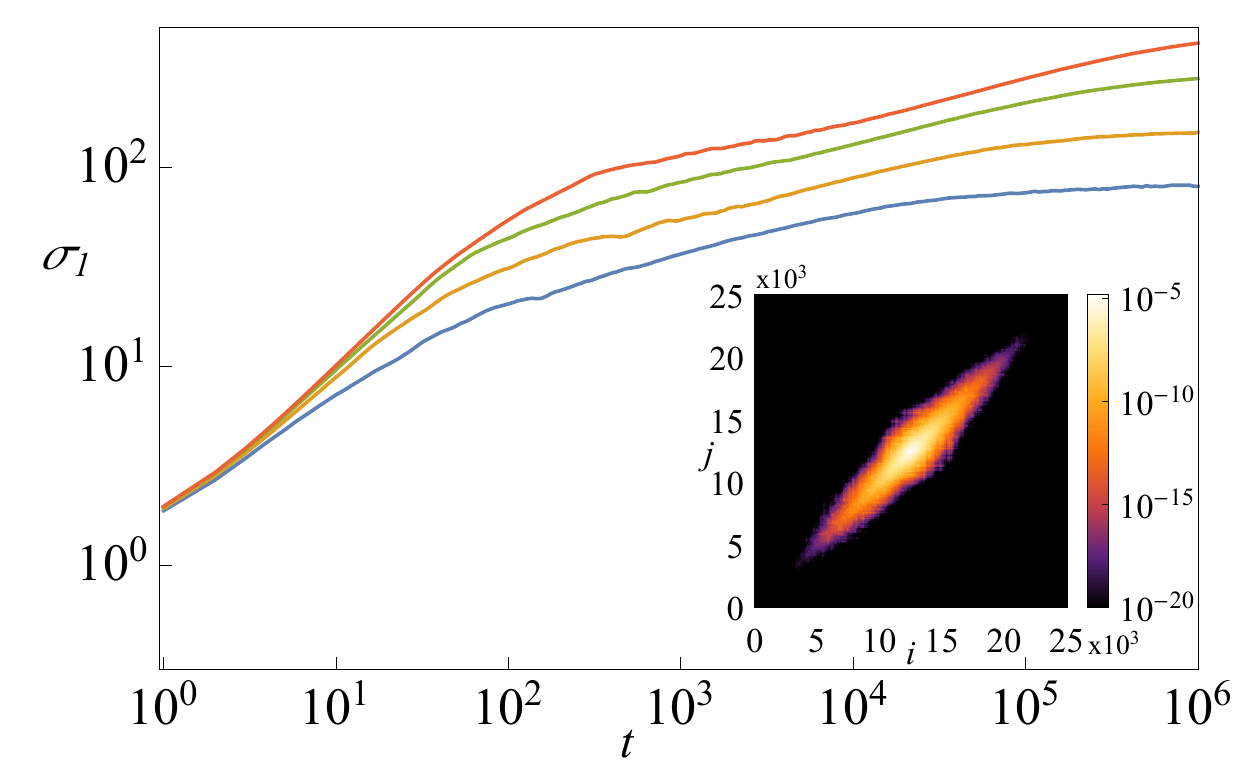}
\caption{\label{fig3} Time evolution of $\sigma_1$
of a TIW for different values of $\theta=\pi/8,\pi/12,\pi/16,\pi/20$ from bottom to top.
Here $\gamma=\pi$ and $N=25000$. 
Inset: snapshot of the probability distribution $p_{ij}$ for $\theta=\pi/20$ at $t=10^6$, showing strongly anisotropic wave packet spreading.
}
\end{figure}

Let us present the numerical analysis of the dynamics of the TIW for  non-vanishing interaction $\gamma \neq 0$. 
The largest absolute value of the term $(e^{i\gamma} -1)$ in (\ref{tip_interaction}) is obtained for $\gamma=\pi$, which we choose as our operational value in this section.
We evolve a system of size $N=25000$ up to time $t_\text{max}=10^6$. We follow the time dependence of the standard deviation $\sigma_1$ for various values of the angle $\theta$. These results are presented in Fig. \ref{fig3} (solid lines). 
$\sigma_1(t)$ shows ballistic-like growth  
($\sigma \propto t$) up to $\sigma_1 \sim \xi_1$ in analogy to the noninteracting case. During this first part of the dynamics, the wave packet spreads up to a length scale
of the order of the single particle localization length $\xi_1$. At variance to the noninteracting case, the interacting dynamics continues beyond the limits set by
the single particle DTQW Anderson localization. The corresponding growth of $\sigma_1$ with time is close to a sub-diffusive one 
$\sigma \propto t^\alpha$ with $\alpha \leq 0.5$. 

For $\theta=\pi/8$ and $\xi_1\approx 12$ we observe saturation of $\sigma_1(t)$ at the largest computational time $t=10^6$. For smaller values of $\theta$ and correspondingly for
larger values of $\xi_1$,  the saturation is shifted to larger time and spatial scales, and becomes barely visible for $\theta=\pi/20$ and $\xi_1 \approx 81$.
Choosing larger system sizes, despite being necessary, turns hard due to CPU time and memory limitations. For practical purposes we therefore will
present data which correspond to the largest evolution times.

\begin{figure}
\includegraphics[width=0.5\textwidth]{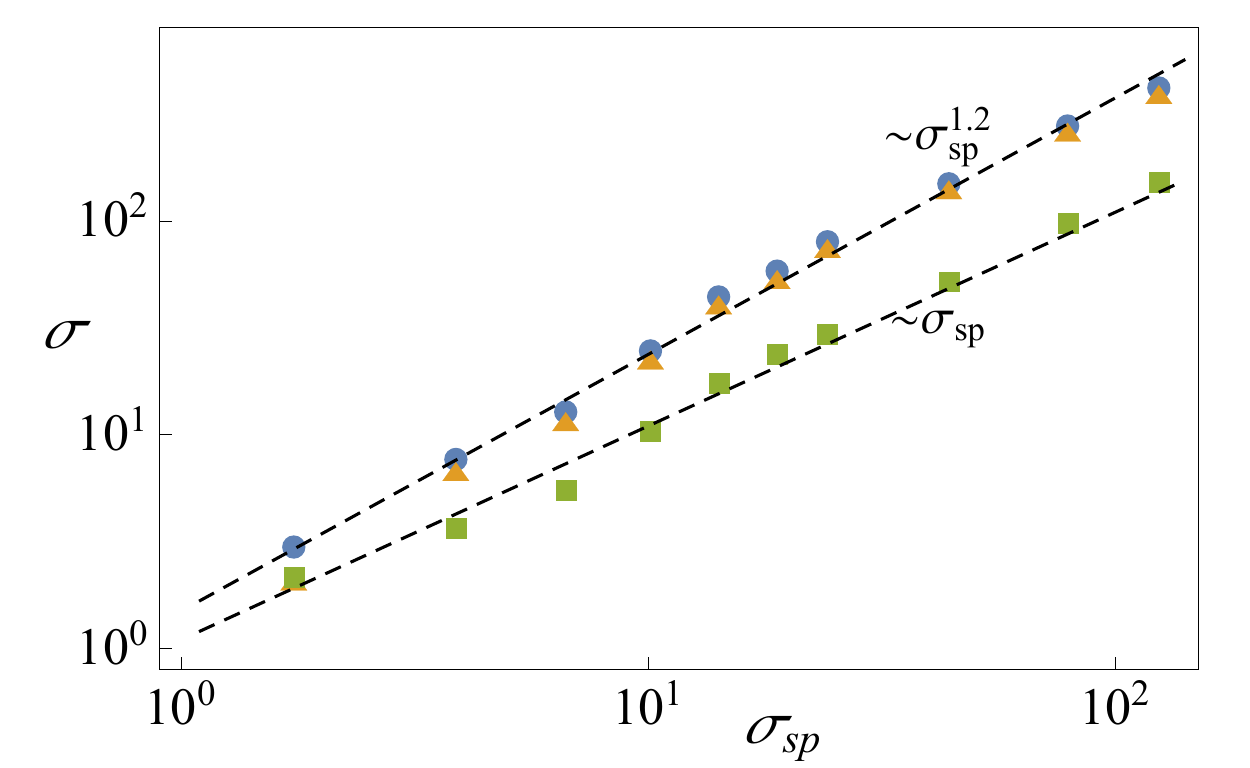}
\caption{\label{fig4} Scaling of the TIW length scale $\sigma_{\perp}$ (green squares), $\sigma_{\parallel}$ (orange triangles) and $\sigma_1$ (blue circles)  with the  single particle DTQW
length scale $\sigma_{sp}$. The corresponding values of $\theta$ vary between $\pi/20$ and  $\pi/3$. 
Here $\gamma=\pi$ and $N=25000$.  
Black dashed lines are algebraic fits. 
}
\end{figure}

In the inset of Fig.\ref{fig3} we plot the probability distribution of wave function  $p_{ij}(t =10^6)$  for $\theta=\pi/20$. It shows a clear reduction to
the two-fold rotational symmetry which leads to the emergence of at least two different length scales $\sigma_{\perp}$ and $\sigma_{\parallel} \gg \sigma_{\perp}$
which characterize the width and elongation of the cigar-like shape.
The dependence of the new length scales on the single particle $\sigma_{sp}$ one is shown in 
Fig.\ref{fig4}. The width $\sigma_{\perp} \approx \sigma_{sp}$ demonstrates that the limit of relative distance on which the two single particle DTQW components of the TIW can propagate is set by $\sigma_{sp}$. However, the elongation $\sigma_{\parallel}$  shows a faster than linear growth with
$\sigma_{sp}$. A simple power law fit $\sigma_{\parallel} \approx \sigma_{sp}^{\beta}$ yields $\beta \approx 1.2$.

\subsection{Varying $\gamma$}

\begin{figure}
\includegraphics[width=0.5\textwidth]{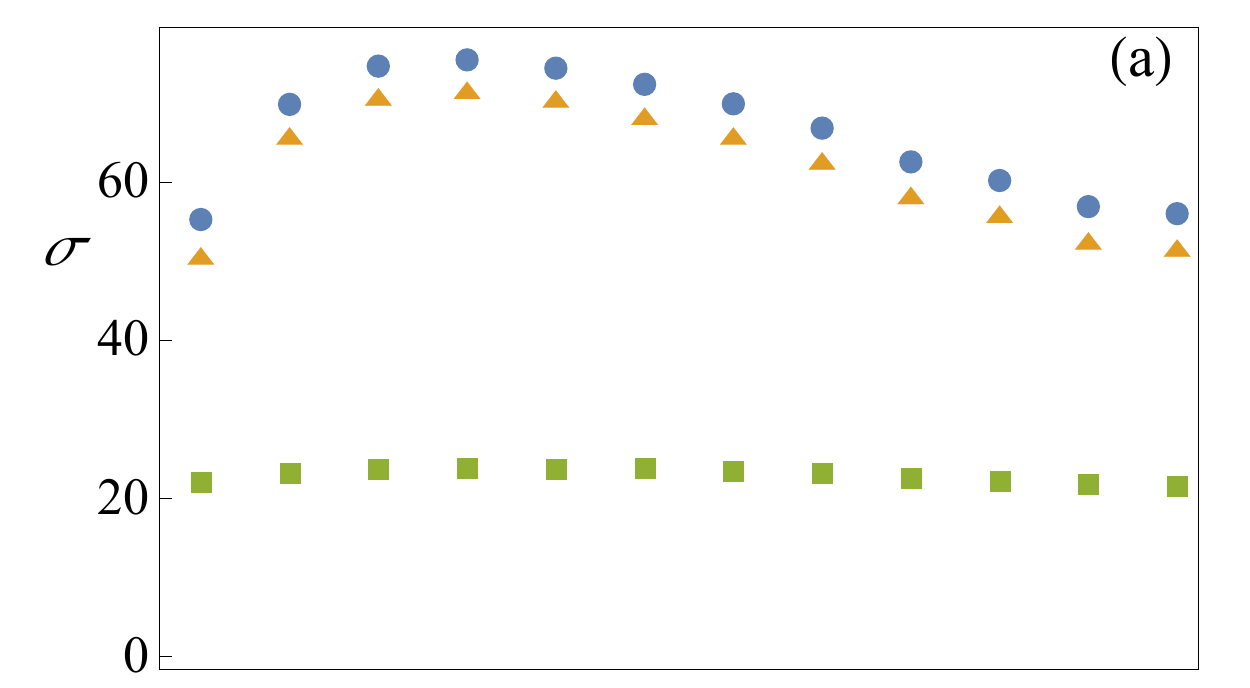}
\includegraphics[width=0.5\textwidth]{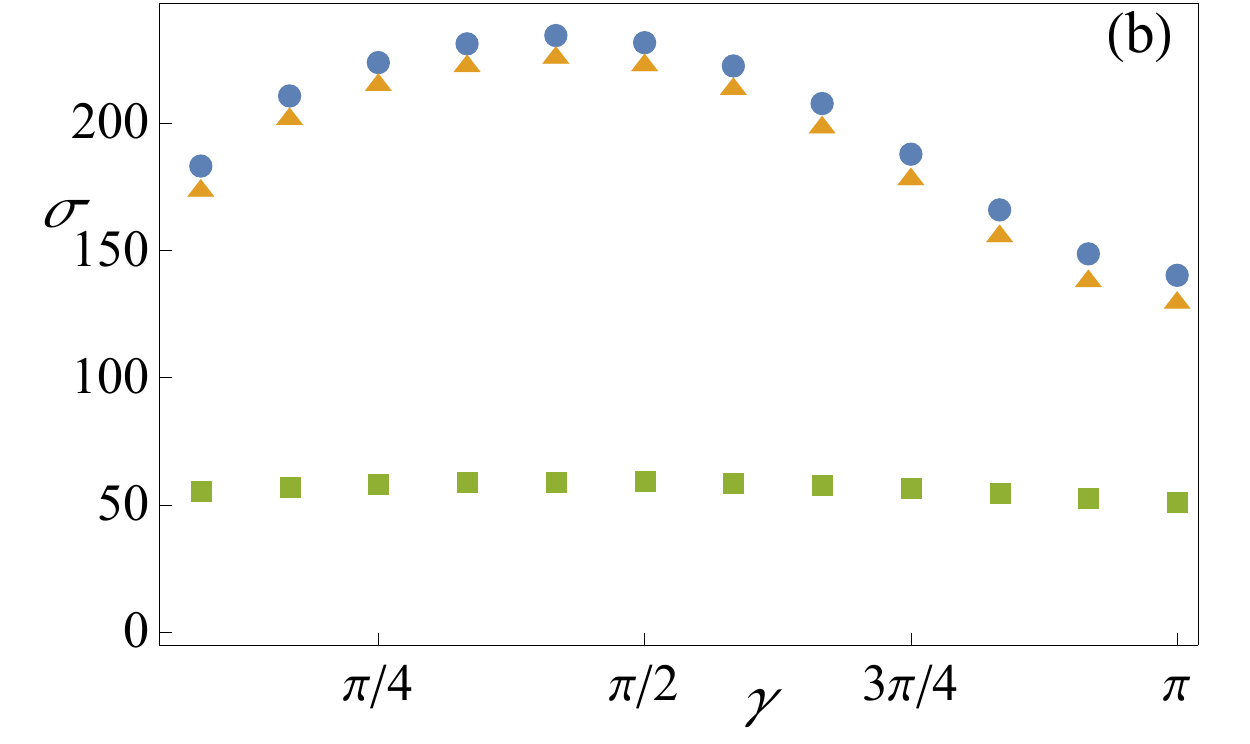}
\caption{\label{fig5} $\sigma_1$ (blue circles), $\sigma_{\parallel}$ (orange triangles), $\sigma_{\perp}$ (green squares) as functions of the interaction parameter $\gamma$.  
(a) $\theta=\pi/8$. (b) $\theta=\pi/12$. 
}
\end{figure}

Finally we study the impact of varying the interaction strength $\gamma$ for two different values of $\theta=\pi/8$ and $\theta=\pi/12$ in Fig.\ref{fig5}.
In order to avoid disorder realization induced fluctuations, we evolve the wave packet up to $t=2 \cdot 10^5$  (which is sufficient for the chosen $\theta$ values) and average over
10 disorder realizations.
We first discuss the data for the width $\sigma_{\perp}$. Since we concluded that $\sigma_{\perp} \approx \sigma_{sp}$ is a single particle DTQW length scale,
it should not depend on the strength of $\gamma$. Indeed, the computational data demonstrate this very clearly.
At the same time, the elongation scale $\sigma_{\parallel}$ respectively $\sigma_1$ should strongly depend on $\gamma$. Again, the computational data in Fig.\ref{fig5}
demonstrate this very clearly. The curves $\sigma_{\parallel ,1}(\theta)$ show a clear maximum at $\gamma_m(\theta)$. Surprisingly, $\gamma_m \neq \pi$, with a weak but observable dependence
on $\theta$. Therefore the value $\gamma=\pi$ is in general not corresponding to the case of strongest enhancement of the TIW localization length. Possibly there is a hidden symmetry in the
TIW problem at $\gamma=\pi$ whose violation for $\gamma\neq \pi$ might lead to an enhancement of the localization length.

\section{Conclusion and outlook}

We analyzed the interplay of disorder and interaction in the Floquet Anderson localization problem of two interacting discrete time quantum walks. 
The single particle DTQW is described by a Floquet unitary map 
defined on a chain of two-level systems. Despite the action of strong disorder in one of the Floquet unitary map parameters, the resulting  novel Anderson localization phase is
characterized by a gapless Floquet spectrum and one unique localization length $\xi_1$
for all single particle eigenstates. The ratio of the participation number of the eigenstates $P$ over $\xi_1$ is not constant, indicating a violation of the usually expected
single parameter scaling regime as known for Hamiltonian disordered systems.
We add a local contact interaction, which is parametrized by a phase shift $\gamma$.
A wave packet is spreading subdiffusively beyond the bounds set by $\xi_1$ and saturates at a new length scale $\xi_2 \gg \xi_1$.
For the assumed strongest interaction case $\gamma=\pi$ we identify a new length scale $\xi_2 \gg \xi_1$ which follows
$\xi_2 \sim \xi_1^{1.2}$. 
We observe a nontrivial dependence of $\xi_2$ on $\gamma$, with a maximum value
observed for $\gamma$-values which are shifted away from the expected strongest interaction case $\gamma=\pi$. We currently lack an understanding of this intriguing fact,
which has to be addressed in future work.
In the absence of interaction $\gamma=0$ we confirm the persistence of the violation of the single parameter scaling. The explanation of this surprising observation
is another interesting topic to be addressed in future work.

\textbf{Acknowledgement.} This work was supported by the Institute for Basic Science, Project Code (IBS-R024-D1).

\end{document}